# ENTANGLEMENT AND SELF-PULSING INSTABILITY

N. H. Adamyan[2], S. B. Manvelyan[1], and G. Yu. Kryuchkyan[1,2]

[1]*Institute for Physical Research, National Academy of Sciences, Ashtarak-2, 378410, Armenia,*
[2]*Yerevan State University, A. Manookyan 1, 375049, Yerevan, Armenia.*

*We study the Wigner function of phase-locked nondegenerate optical parametric oscillator and find the signatures of both phase-locking and self-pulsing phenomena in phase space. We also analyze the problem of continuous-variable entanglement in the self-pulsing instability regime.*

**Introduction**

In the presently very active field of continuous variable (CV) quantum information processing [1] a challenging goal consists in generation of entangled intensive light beams. Various quantum optical schemes generating entangled bright light have been proposed for this goal. However, up to now the generation of bright light states with high level of CV entanglement meets serious problems. For instance, type-II nondegenerate optical parametric oscillator (NOPO) can generate such quantum states. It is well known that the relative phase between subharmonics in above threshold NOPO undergoes diffusion process. This process destroys the frequency degeneracy of modes and hence limits the production of stable quantum-twin beams and CV entanglement in NOPO above threshold. A more stable type II NOPO, operated under modes phase-locked condition was first demonstrated in [2]. Quantum theory of this device that contained an intracavity quarter-wave plate for coupling two orthogonally polarized modes of the subharmonics has been recently studied in detail and CV entangled states of light beams in the presence of phase localizing process has been proposed in [3]. The theoretical and experimental investigations of CV entanglement with self-phase-locked type II OPO have been presented in [4]. Note that all these studies have been performed in steady-state stable regime of NOPO. Recently, Hopf instabilities in a triply resonant OPO have been demonstrated in [5].

In this paper we continue investigation of phase-locked NOPO follow the paper [3] and consider also the nonstationary regime of generation. It will be shown that the system of interest is characterized by the self-pulsing instability which connects the steady-state regime of mode generation below threshold, to a temporal periodic regime. Our goal is two-fold. In one part of the present paper, we expand the previous study of stable, stationary phase-locked NOPO [3], considering the important details in phase-space with help of the quasiprobability distribution function (Wigner function). The other part of the paper is devoted to the problem of entanglement in self-pulsing instability regime.

**Results**

The system under consideration is the NOPO with quarter wave plate inside the cavity, interacting with the thermal bath. Due to explicit presence of dissipation in this problem, one has to write the master equation for the reduced density matrix of the system, which within the framework of the rotating wave approximation and in the interaction picture is

$$\frac{\partial \rho}{\partial t} = \frac{1}{i\hbar}[H,\rho] + \sum_{i=1}^{3} \gamma_i \left(2a_i \rho a_i^+ - a_i^+ a_i \rho - \rho a_i^+ a_i \right), \qquad (1)$$

where

$$H = \sum_{i=1}^{3} \hbar \Delta_i a_i^+ a_i + i\hbar E\left(a_3^+ - a_3\right) + i\hbar k\left(a_3 a_1^+ a_2^+ - a_3^+ a_1 a_2\right) + \hbar\chi\left(a_1^+ a_2 + a_1 a_2^+\right), \qquad (2)$$

and $a_i$ are the boson operators for the cavity modes $\omega_i$. The mode $a_3$ at frequency $\omega$ is driven by an external field with amplitude $E$, while $a_1$ and $a_2$ describe subharmonics of two orthogonal polarizations at degenerate frequencies $\omega/2$ generated in the process $\omega \to \frac{\omega}{2} + \frac{\omega}{2}$. The constant $k$ determines the efficiency of the down-conversion process, while $\chi$ describes the energy exchange between only the subharmonic modes due to the intracavity waveplate. We take into account the detunings of subharmonics $\Delta_i$ and the cavity damping rates $\gamma_i$ and consider the case of high cavity losses for pump mode ($\gamma_3 << \gamma_1, \gamma_2$), when this mode can be adiabatically eliminated. However, in our analysis we take into account the pump depletion effects. We solve Eq.(1) by using the well known numerical quantum state diffusion method (QSD) according to which the density matrix is described as an ensemble averaging over quantum trajectories [6].

The semiclassical equations of self-locked NOPO for the complex c-number variables $\alpha_i$ and $\beta_i$ corresponding to the operators $a_i$ and $a_i^+$, have the following form [3]:

$$\frac{\partial \alpha_1}{\partial t} = -(\gamma_i + i\Delta_1)\alpha_1 + (\varepsilon - \lambda\alpha_1\alpha_2)\beta_2 - i\chi\alpha_2, \qquad (3)$$

$$\frac{\partial \beta_1}{\partial t} = -(\gamma_i - i\Delta_1)\beta_1 + (\varepsilon - \lambda\beta_1\beta_2)\alpha_2 + i\chi\beta_2. \qquad (4)$$

Here $\varepsilon = kE/\gamma_3$, $\lambda = k^2/\gamma_3$, $\beta = \alpha^*$. The equations for $\alpha_2$ and $\beta_2$ are obtained by changing the indexes $1 \leftrightarrow 2$.

It has been shown in [3] that these equations has stationary solutions only if the following relation holds

$$4\chi^2\Delta_1\Delta_2 > (\gamma_1\Delta_2 - \gamma_2\Delta_1). \qquad (5)$$

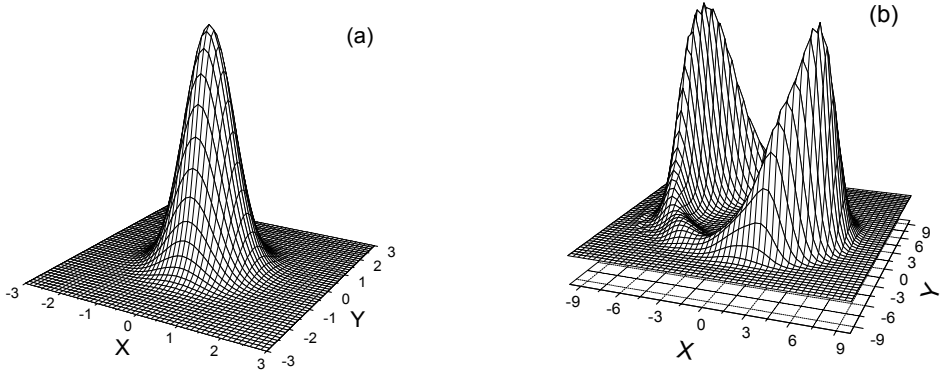

*Fig.1.* The Wigner function of the self-phase locked NOPO in stationary regime for the parameters: $\lambda/\gamma = 0.1$, $\Delta_1/\gamma = \Delta_2/\gamma = 10$, $\chi/\gamma = 0.1$, (a) $\varepsilon/\gamma = 5$, (b) $\varepsilon/\gamma = 11$.

At first let us turn to the stationary regime and consider the phase locking phenomenon in framework of the Wigner function. In Fig. 1 the Wigner function of one of the subharmonics of self locked NOPO for (a) below, and (b) above-threshold are presented for over transient regime. In above-threshold regime the Wigner function has two peaks (which acquire crescent form) due to the phase locking of the subharmonic modes. The distance between the peaks increases with increasing of the photon number (decreasing $\lambda/\gamma$). The increasing of the wave plate parameter $\chi$ leads to increasing of the phase-locking, i.e. to more localization of the peaks.

Now let us pay our attention to the case of nonstationary regime of generation, when the inequality (5) does not valid. We find that the classical solution exhibits the self-pulsing instability: the photon number of intracavity modes oscillates periodically. We demonstrate this oscillations in Fig.2 (a) (curve 1) for one of the modes. It appears that self-pulsing exists in the whole range of the inequality (5), and does not depend on what parameter is changed to violate it. An individual quantum trajectory of the single mode of self-locked NOPO for the same parameters is presented in Fig.2 (b). It is seen that trajectory roughly repeats the oscillations of semiclassical solution. Deviations are due to stochasticity of a single trajectory. The result, averaged over ensemble of quantum trajectories is shown in Fig. 2 (a) (curve 2).

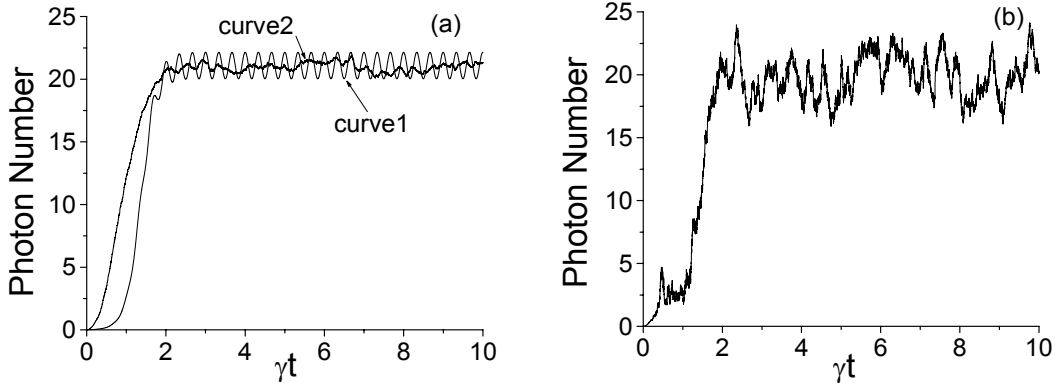

***Fig.2.*** *(a) Classical trajectory (curve 1) and quantum ensemble averaged result (curve 2), and (b) quantum trajectory of self-phase locked NOPO in the regime of self-pulsing. The parameters are:* $\lambda/\gamma = 0.1, \Delta_1/\gamma = 10, \Delta_2/\gamma = -5, \chi/\gamma = 0.1, \varepsilon/\gamma = 4$.

The instability clearly manifests itself in the phase space of the quantum system. Indeed, the Wigner function in this case has an elliptical form (see Fig.3 (b)), which is the reflection of classical phase space trajectory (see Fig.3 (a)). (This property of Wigner function to repeat the shape of classical phase trajectory has been discussed in several papers and also is concerned to the problem of quantum chaos (see [7] and reference therein)). The elliptical form of the Wigner function indicates that the self-phase locked NOPO also suffer phase diffusion. Indeed, a phase in the phase space is described as a polar coordinate. As the Wigner function distributed almost uniformly to all directions, this mean that the phase operator is entirely uncertain, that is equivalent to a phase diffusion.

To characterize the CV entanglement in the self locked NOPO we use the inseparability criterion for the quantum state of two optical modes: $V = \frac{1}{2}(V_+ + V_-) < 1$, where $V_+$ and $V_-$ are the relevant distance and total momentum of the system correspondingly (see [3] for details). For investigation of interplay between entanglement and self-pulsing, we present in Fig.4 the dependence of $V$ on $\varepsilon/\varepsilon_{th}$, (where $\varepsilon_{th}$ is the threshold value of $\varepsilon$) for the nonstationary regime. This variance has been found in [3] for the stationary regime, and the case of the equal detunings when $\varepsilon_{th} = \sqrt{(\chi - |\Delta|)^2 + \gamma^2}$, while in the self-pulsing regime it is complicated to find an analogous analytical result. So, we estimate $\varepsilon_{th}$ for the latter case numerically and obtain: $\varepsilon_{th}/\gamma = 1$ for the parameters of Fig.4. It is seen from Fig.4 that CV entanglemen is realized for wide ranges of the ratio $\varepsilon/\varepsilon_{th}$. The variance drops down near the threshold and achieves minimal value $V = 0.55$ at the threshold.

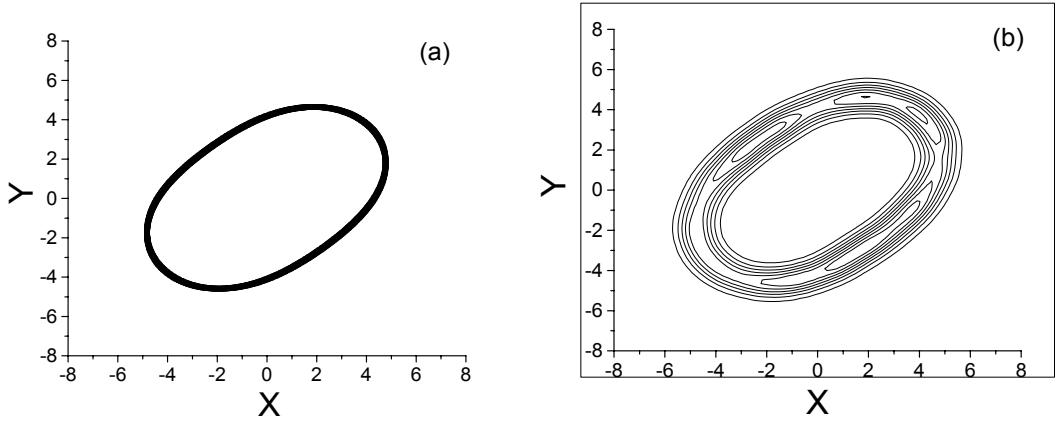

*Fig.3.* *Classical phase space trajectory (a) and contour-plot of Wigner function (b) of the self--phase locked NOPO in the regime of self pulsing. The parameters are:* $\lambda/\gamma = 0.1, \Delta_1/\gamma = 0.1, \Delta_2/\gamma = -0.1, \chi/\gamma = 0.5, \varepsilon/\gamma = 3$.

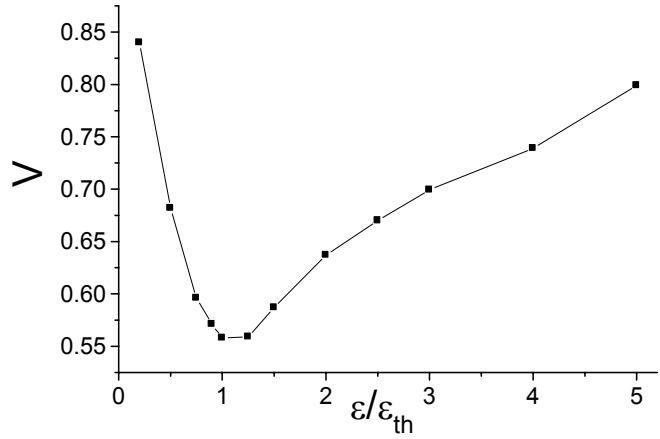

*Fig.4.* *The variance V versus pump amplitude for the nonstationary regime. The parameters are:* $\lambda/\gamma = 0.1, \Delta_1/\gamma = 10, \Delta_2/\gamma = -10, \chi/\gamma = 0.1$.

In conlusion, we have reported on phase-space aspects of phase-locked NOPO. Future investigations of this subject may lead to a contribution in physics of entanglement of ultrastable light beams.

**Acknowledgement**

This work was supported by the NFSAT PH 098-02/CRDF 12052 grant.

**References**

[1] Quantum Information Theory with Continuous Variables, S.L. Braunstein and A.K. Pati, eds., Kluwer, Dordrecht, (2003).
[2] E.J. Manson and N.C. Wong, Opt. Lett., **23**, 1733 (1998).
[3] H.H. Adamyan, G.Yu. Kryuchkyan, Phys.Rev. A **69**, 053814 (2004).
[4] L.Longchambon, et. al., Eur. Phys. J. D **30**, 287 (2004); J. Laurat et. al., Phys.Rev. A **70**, 042315 (2004).
[5] J-J. Zondy et. al., Phys.Rev. Lett. **93**, 043902 (2004).
[6] Quantum State Diffusion, I.C. Percival, Cambrige University Press (2000).
[7] G.Yu. Kryuchkyan and S. B. Manvelyan, Phys.Rev. Lett. A **68**, 013823 (2003).